# Structural and phase evolution in $U_3Si_2$ during steam corrosion

Jiatu Liu[1], Patrick A. Burr[1], Joshua T. White[5], Vanessa K. Peterson[2,3], Pranesh Dayal[2], Christopher Baldwin[2], Deborah Wakeham[2], Daniel J. Gregg[2], Elizabeth S. Sooby[4], Edward G. Obbard*[1]

1, School of Mechanical and Manufacturing Engineering, UNSW Sydney, NSW 2052, Australia

2, Australian Nuclear Science and Technology Organisation, Locked Bag 2001, Kirrawee DC, NSW 2232, Australia

3, School of Mechanical, Materials, and Mechatronic Engineering, University of Wollongong, Wollongong, NSW 2522, Australia

4, Department of Physics and Astronomy, The University of Texas at San Antonio, San Antonio, TX 78249, USA

5, Material Science and Technology Division, Los Alamos National Laboratory, Los Alamos, NM, USA

*Corresponding author: e.obbard@unsw.edu.au

## Abstract

$U_3Si_2$ nuclear fuel is corroded in deuterated steam with in situ neutron diffraction. Density functional theory is coupled with rigorous thermodynamic description of the hydride including gas/solid entropy contributions. H absorbs in the 2*b* interstitial site of $U_3Si_2H_x$ and moves to 8*j* for $x \geq 0.5$. Hydriding forces lattice expansion and change in a/c ratio linked to site preference. Rietveld refinement tracks the corrosion reactions at 350-500 °C and preference for the 8*j* site. Above 375 °C, formation of $UO_2$, $U_3Si_5$ and $USi_3$ take place in the grain boundaries and bulk. Hydriding occurs in bulk and precedes other reactions.

## Introduction

Following the Fukushima Daiichi accident, nuclear fuel industry and researchers revisited material selection for power reactor fuel and cladding systems. The aim was to prevent or delay the type of core damage that results from station wide blackout and loss of post shutdown core cooling. This class of fuels is known as accident tolerant fuels (ATF). One such candidate, $U_3Si_2$, has a higher uranium density (11.3 g/cm$^3$) and high thermal conductivity (about 15 W/mK at 600 °C) than the widely used $UO_2$ [1]. $U_3Si_2$ is also applied as a research reactor fuel, in the form of $U_3Si_2$ dispersion encapsulated in aluminum plate [2] and fuel of the Australian OPAL reactor is one example of this [3].

If deployed in a power reactor, the attractive physical properties of $U_3Si_2$ compared to conventional $UO_2$ mean that fuel pins could operate with lower centerline temperature. The reduction in stored thermal energy, decreases pumped cooling requirements and/or increases "grace time" before core damage in some accident scenarios [4]. The high density and higher uranium loading of $U_3Si_2$ would enable exchange of zirconium cladding with more oxidation resistant alloys, such as oxide dispersion strengthened steels or SiC/SiC, or addition of protective coatings [5], without compromising neutronic performance of the reactor.

The physical and thermal properties of $U_3Si_2$ in relation to its application as power reactor fuel have been investigated over the past decade [5], including the in situ measurement of thermal expansion in $U_3Si_2$ to 1600 °C by neutron diffraction experiments [6,7]. $U_3Si_2$ has been assessed for fission gas swelling [8] and phase stability [9], with generally satisfactory results in relation to its envisaged use as ATF.

Sooby Wood *et al.* investigated the corrosion behavior of $U_3Si_2$ in steam by thermogravimetry (TG), finding a rapid pulverizing reaction at elevated temperatures (above 350 °C) and possible formation of a silicon rich phase $USi_3$ by ex situ X-ray diffraction (XRD) analysis [10]. TG in this case did not enable reliable analysis of the corrosion process as it occurred because the energetic nature of the hydriding/oxidation reaction expelled pulverized materials from the TG crucible. From microstructural characterization and comparison between corrosion in steam and hydrogen, volumetric expansion of an intermediate $U_3Si_2$-hydride phase was identified as a probable reason for the pulverization effect [10].

This hydride has the same space group, P4/mbm, as that of $U_3Si_2$ [11]. Orthographic views of $U_3Si_2$ with possible H interstitial sites are shown in Figure 1. The H stoichiometry can be as high as $U_3Si_2H_{1.8}$ with 10% volume expansion when $U_3Si_2$ is exposed to 10 MPa $H_2$. Mašková *et al.* originally proposed the 8*j* site for H, but could not confirm this experimentally because the X-ray scattering cross section of H is negligible compared with those of U and Si. Middleburgh *et al.* and Shivprasad *et al.* calculated the solution energies of H at different crystallographic positions in a unit cell of $U_3Si_2$, for stoichiometry ranging $U_6Si_4H_{1-4}$ [11,12]. They reinforced 8*j* site as the most favorable for H in this small, and therefore highly ordered, cell and without considering thermal and entropic effects.

The present work researches the role and the structure of the hydride phase in this form of corrosion, using a combination of in situ neutron diffraction, microscopy and DFT simulations. We use steam at ambient pressure and isothermal temperatures between 350-500 °C similar to [10], and track the corrosion process by phase analysis and crystal structure. We observe in real time the formation of hydride phase $U_3Si_2H_x$, plus $USi_3$, $U_3Si_5$ and $UO_2$ phases. DFT calculation of free energy and lattice parameters for several H positions and concentrations that include ensemble averaged formation energies and configurational entropies are compared to diffraction results for H position and lattice parameters in $U_3Si_2$. Microstructures of the corroded samples are analyzed after the in situ experiment by XRD, scanning electron microscopy and energy dispersive spectroscopy (SEM/EDS).

## Material and methods

Specimens of $U_3Si_2$ were prepared by arc-melting ingots from metal U and Si followed by powder metallurgical sintering in a high temperature furnace [7]. Ingots were arc melted in a high purity Ar maintained below 20 wppm O. Powder comminution for powder metallurgy was accomplished in a high energy ball mill prior to pressing and loading into a W-mesh element furnace. All milling, pressing, and furnace loading operations were conducted in a high purity Ar glove box line to minimize oxidation of the powder. Sintered pellet densities were determined geometrically between 88-96% of the theoretical density of $U_3Si_2$.

In situ diffraction data were collected on Wombat high intensity powder diffractometer at the Australia Centre for Neutron Scattering (ACNS) [13]. Constant wavelength neutron diffraction patterns were collected continuously for 1 minute on the 2-dimensional detector, corrected and reduced to one dimensional powder diffraction histograms. The wavelength and instrumental parameters were refined against NIST La$^{11}$B$_6$ 660b standard reference material [14]. A 316L stainless-steel sample tube was designed to confine the sample in the neutron beam, allowing delivery of deuterated steam while inside a ILL-type vacuum furnace. The sample environment was controlled by a Hidden Isochema XCS system.

The first sample, $U_3Si_2\_1$, was heated to 350 °C at 10 °C/min and held for 6 hours, followed by cooling. Then it was heated to 375 °C at 10 °C/min and held for 32 min before ramping to 400 °C at the same rate and holding for 66 min, followed by cooling. $D_2O$ steam was delivered at 3.5 g/h by He flow of 0.5 L/min. Diffraction patterns were recorded continuously except for the first cool-down period.

A second sample, $U_3Si_2\_2$, was heated to 500 °C at 4 °C/min and held for 10 min before cooling. $D_2O$ was delivered as for the first sample, but halted unavoidably 10 min after the start of cooling.

Post experiment XRD was performed on the sample residues without crushing, in a Bruker D8 Advance diffractometer using Cu Kα radiation ($\lambda_1$=1.5405 Å, $\lambda_2$=1.5443 Å) and a Lynxeye detector in the 2θ range 10-80° with a step size of 0.02° for 10 hours each. XRD peak profiles were determined from alumina standard reference material from Bruker.

SEM/EDS used a Tescan Fera 3 SEM instrument with an attached EDS detector (ThermoFisher Scientific) at an accelerating voltage of 15 kV. For topographic imaging, the fragmented sample residue was sprinkled on carbon-based, electrically conductive, double sided adhesive discs on SEM specimen stubs and coated with ~10 nm of carbon film. For cross sections, relatively large pieces of the fragmented residual material were mounted in epoxy resin and then polished to a 1 μm-diamond finish and coated with ~10 nm of carbon film. Pathfinder$^{TM}$ X-ray microanalysis software was used for EDS data acquisition and standardless EDS data analyses.

Rietveld and other analysis [15] was used to identify the hydrogen position/occupancy within the material and track the structural evolution during the in situ corrosion experiment. GSAS software [16] together with gsaslanguage [17] were used to perform refinements.

Using deuterated steam, the comparable coherent scattering cross sections of the constituent elements ($^2$H 5.59, Si 2.16, O 4.23, U 8.9 barn) enable Rietveld refinement of hydrogen position and occupancy of $^2$H in the $U_3Si_2$. However, the data in this case did not permit unconstrained refinement. We systematically defined a series of prospective crystallographic sites for the $^2$H and performed refinement for each of these against a single diffraction pattern that showed hydrogen uptake, indicated by a clear change in the $U_3Si_2$ lattice parameters. $^2$H occupancy values equivalent to the stoichiometries $U_3Si_2H_{0.4}$, $U_3Si_2H_{0.8}$, $U_3Si_2H_{1.2}$, $U_3Si_2H_{1.6}$, and $U_3Si_2H_2$ were also tested. We define the 'best' structural model as that which gives the smallest product $\chi^2 \cdot R_{F^2}$ of two fitting criteria [16].

The prospective hydrogen positions were intentionally independent of the DFT work – being the top 12 ranked in terms of maximum interatomic distance to neighboring U or Si atoms in room temperature $U_3Si_2$. This resulted in selection of positions with U-H distances larger than 210 pm and Si-H distances larger than 142 pm. These values are the sum of individual U-H or Si-H atomic radii [18] with a 5% surplus. According to symmetry, we checked nearly 35,000 candidate positions in 1/16 of the unit cell volume shown in Figure 1 with step size 0.2% *a*, 0.2% *b*, 0.1% *c*. All candidate positions are in polyhedra, such as trigonal bipyramidal 3U-2Si, tetrahedral 4U, 2U-2Si, and 3U-Si. The longest U-H distance found is 2.249 Å, less than that in α-$UH_3$ (2.32 Å) [19] or β-$UH_3$ (2.26 Å) [20], while the longest Si-H distance 2.906 Å, much larger than that in $SiH_4$ (1.48 Å) [21]. The shortlisting process was judged to be exhaustive. Locally maximum U-H distance within a surrounding volume of 0.4% *a* x 0.4% *b* x 0.3% *c* provided 12 candidate positions shown in Figure 1. These H positions, combined with the 5 occupancy values were used in Rietveld models.

Sequential refinement across all temperature points refined lattice parameters and phase fractions of all phases, assuming the best $U_3Si_2H_x$ structure defined previously. For both $U_3Si_2\_1$ and $U_3Si_2\_2$, $U_3Si_2H_x$ ($P4/mbm$) and $UO_2$ ($Fm\overline{3}m$) phases were used to refine against diffraction data until the refinement became unstable. Then $U_3Si_5$ ($P6/mmm$) and $USi_3$ ($Pm\overline{3}m$) [22] were added into the refinement. $U_3Si_2H_x$ was removed from the Rietveld model when the phase fraction became negligible. With only one discernable peak, lattice parameters $a$ and $c$ of $U_3Si_5$ phase were constrained equal.

Sequential refinement involved a degree of assumption for the hydrogen occupancy. For $U_3Si_2\_1$, occupancy at $8j$ site was fixed at 0.3 ($U_3Si_2H_{1.2}$) for all refinements during 375 °C holding and the preceding temperature ramping period. H occupancy of 0.1 ($U_3Si_2H_{0.4}$) was used for all refinements during 350 °C holding. $U_3Si_2$ phase without H was used for the temperature ramping to 350 °C. For sample $U_3Si_2\_2$ a $U_3Si_2$ phase without H was adequate for sequential refinements.

## Calculation

DFT calculations were performed to investigate the thermodynamics and structure of $U_3Si_2H_x$. Two sets were performed to investigate dilute and concentrated H levels. Calculations were performed within the program VASP [25,26], with the PBE exchange correlation functional [27], pseudopotentials from the VASP 5.2 library containing 14, 4 and 1 valence electrons for U, Si and H atoms respectively, and a planewave energy cut-off energy of 350 eV. All simulations were spin-polarized. An effective on-site coulomb repulsion term of 1.5 eV was applied to the $f$ electrons of U atoms, using the rotationally invariant formalism of Dudarev et al [28], as this was shown to improve the description of $U_3Si_2$ in previous work [29]. The need for the on-site coulomb correction was confirmed here by repeating a large subset of the calculation without the correction, and these simulations exhibited unstable structures with substantial relaxation and reconfigurations (Table S3).

To represent dilute H solution, a single H atom was placed in a 2x2x4 supercell of $U_3Si_2$ (160 atoms). H atoms were inserted in the 16 unique interstitial sites. For comparison with previous work, these were repeated in a conventional unit cell, and in a 1x1x2 supercell, which was later used to investigate solutions with high H concentration. Both atomic position and lattice vectors were relaxed during energy minimization. For the largest cell, the degrees of freedom were reduced by constraining the internal coordinates of atoms further than 3.5 Å from the H atom to prevent unphysical collapse of the cell that was occasionally observed. This constraint introduces an artifact energy that is known to be negligible in the case of a small and charge-neutral defect such as a hydrogen interstice in $U_3Si_2$ [30]. Interestingly, not using the on-site coulomb correction resulted in sizable distortion of the structure even for an defect-free crystal, and irrespective of supercell size.

Solid solutions of $U_3Si_2H_x$ with high hydrogen concentrations ($0.5 \leq x \leq 2$) were modelled considering all interstitial sites that proved stable under dilute conditions, and sufficiently far from one another. When low symmetry sites were too close to one another to conceivably fit multiple H interstitial atoms, the central higher-symmetry site was considered instead. For example, the $2b$ site is surrounded by eight $16l$ sites, each only 0.28 Å from the central $2b$ site. These small displacements from the central site (in any one of eight directions) are expected to average during a diffraction experiment, yielding the same average position and slightly increase $U_{iso}$. The concentrated solid solutions were modelled using a configurational ensemble [28], which makes no assumptions of order/disorder, instead it relies on explicit calculations of all symmetrically unique H configurations for a given composition to extract the exact configurational entropy of the system. We used the site occupation disorder (SOD) code to reduce the set of simulations to those that are symmetrically unique [31]. In brief, the ensemble probability of occupying a particular configuration $k$ is calculated as

$$P_k = \frac{\Omega_k}{Z} \exp\left(\frac{-E_k}{k_B T}\right) \tag{1}$$

where $\Omega_k$ is the degeneracy of state $k$ (the number of symmetrically equivalent states) and $Z$ is the partition function

$$Z = \sum_{k=1}^{K} \Omega_k \exp\left(\frac{-E_k}{k_B T}\right) \tag{2}$$

In a configurational ensemble, the effective energy of the system at temperature T is simply

$$E = \sum_{k=1}^{K} P_k E_k \tag{3}$$

This allows to define the configurational entropy in terms of state probabilities:

$$S^{config} = k_B \sum_{k=1}^{K} P_k \ln P_k \tag{4}$$

The contribution of magnetic and electronic entropy is expected to be small in $U_3Si_2H_x$, and near-constant with increasing H content, thus cancelling out in the formation energy. Vibrational entropy, and quasi-harmonic effects could also play a role, but these were not included due to practical limitations.

Physical properties of the ensemble can also be calculated through configurational average. Taking for example the unit cell volume after hydrogen absorption, V:

$$V = \sum_{k=1}^{K} P_k V_k \tag{5}$$

Next, we describe the thermodynamic framework used to find the Gibbs free energy of formation, $\Delta G^f$, of the hydride is defined as

$$\Delta G^f_{U_3Si_2H_x}(T, p_{H_2}) = g_{U_3Si_2H_x}(T) - 3\mu_U(T) - 2\mu_{Si}(T) - x\mu_H(T, p_{H_2}) \tag{6}$$

Where $g_{U_3Si_2H_x}$ is the Gibbs free energy per formula unit of $U_3Si_2H_x$, and $\mu_i$ is the chemical potential of constituent element $i$. We assume the $U_3Si_2H_x$ to be in equilibrium with $U_3Si_2$, thus we take the chemical potentials of U and Si as

$$g_{U_3Si_2} = 3\mu_U - 2\mu_{Si} \tag{7}$$

Note that the temperature dependence has been dropped, as the entropy of thermally activated defects in an ordered solid is negligible compared to the entropy of the gas phase and disordered solids. We can therefore take $g_{U_3Si_2}$ as the DFT energy of a formula unit of $U_3Si_2$:

$$g_{U_3Si_2} = E^{DFT}_{U_3Si_2} \tag{8}$$

while

$$g_{U_3Si_2H_x}(T) = E^{DFT}_{U_3Si_2H_x} - TS^{config}_{U_3Si_2H_x} \tag{9}$$

where $S^{config}_{U_3Si_2H_x}$ is the configurational entropy of the hydride, which is non-negligible due to the disorder on the H sublattice.

The last term in equation (6) is the chemical potential of H, $\mu_H(T, p_{H_2})$, which is described by the law of ideal gasses as

$$\mu_H(T, p_{H_2}) = \mu_H^0(T) + \frac{1}{2} k_B T \log\left(\frac{p_{H_2}}{p_{H_2}^0}\right) \tag{10}$$

where $p_{H_2}$ is the partial pressure of H₂ and a subscript 0 implies standard state. In DFT studies, this quantity is often approximated as half the DFT energy of a H₂ gas molecule, but this approach has two significant limitations: firstly, it is well known that DFT provides a poor description of dimer molecules, and secondly that approach does not account for the entropy of the gas, which is large and cannot be neglected as it is done for solids. To overcome the challenge of modeling a gas molecule at the DFT level of theory, we define $\mu_H$ in terms of the formation energy of a reference hydride, in a similar way that is typically done to define the chemical potential of oxygen by using a reference solid oxide [23]. We consider the following thermodynamic cycle (note that any other thermodynamic cycle could have been selected, provided the relevant thermodynamic quantities are known):

$$\alpha\text{-U}_{(s)} + \frac{3}{2}\text{H}_{2(g)} \rightleftharpoons \beta\text{-UH}_{3(s)} \tag{11}$$

The Gibbs free energy balance of this reaction at standard state is

$$\Delta G^0_{\text{UH}_3}(T) = g^0_{\text{UH}_3}(T) - \mu^0_U(T) - 3\mu^0_H(T) \tag{12}$$

re-arranging for $\mu_H$ and introducing an arbitrary reference temperature $T_0 = 298$ K, we get

$$\mu^0_H(T) = \frac{1}{3}\{g^0_{\text{UH}_3}(T_0) - \mu^0_U(T_0) - \Delta G^0_{\text{UH}_3}(T_0)\} + \Delta\mu^0_H(T) \tag{13}$$

where

$$\Delta\mu^0_H(T) = \mu^0_H(T) - \mu^0_H(T_0) \tag{14}$$

This formulation allows us to calculate $\mu^0_H(T)$ from known or easily-calculatable quantities. Specifically, the absolute energies of solid phases $g^0_{\text{UH}_3}(T_0)$ and $\mu^0_U(T_0)$ can be approximated as the DFT energy of the phase, $E^{DFT}_{\text{UH}_3}$ and $E^{DFT}_{\alpha\text{-U}}$, respectively. While relative quantities $\Delta G^0_{\text{UH}_3}(T_0)$ and $\Delta\mu^0_H(T)$ are readily found in thermodynamic tables. $\Delta\mu^0_H(T)$ is also accurately described by the rigid dumbbell gasses expression (avoiding the need for tabulated experimental values):

$$\Delta\mu^0_H(T) = -\frac{1}{2}\left[(S^0_{H_2} - C^0_P)(T - T^0) + C^0_P T \ln\left(\frac{T}{T^0}\right)\right] \tag{15}$$

where the standard state heat capacity ($C^0_P$) and entropy ($S^0_{H_2}$) of H₂ gas are 0.5998 meV/K and 2.71 meV/K, respectively.

Combining back into equation (10), this yields a complete description of the chemical potential of H:

$$\mu^0_H(T, p_{H_2}) = \frac{1}{3}\{g^0_{\text{UH}_3}(T_0) - \mu^0_U(T_0) - \Delta G^0_{\text{UH}_3}(T_0)\} + \Delta\mu^0_H(T) + \frac{1}{2}k_B T \log\left(\frac{p_{H_2}}{p^0_{H_2}}\right) \tag{16}$$

Altogether, the Gibbs free energy of formation of equation (6) is expanded as

$$\begin{aligned}\Delta G^f_{\text{U}_3\text{Si}_2\text{H}_x}&(T, p_{H_2}) \\ &= E^{DFT}_{\text{U}_3\text{Si}_2\text{H}_x} - TS^{config}_{\text{U}_3\text{Si}_2\text{H}_x} - E^{DFT}_{\text{U}_3\text{Si}_2} - \frac{x}{3}\left[E^{DFT}_{\text{UH}_3} - E^{DFT}_{\alpha\text{-U}} - \Delta G^0_{\text{UH}_3}(T_0)\right] - x\Delta\mu^0_H(T) \\ &\quad - \frac{x}{2}k_B T \log\left(\frac{p_{H_2}}{p^0_{H_2}}\right)\end{aligned} \tag{17}$$

At standard H₂ partial pressure, the last term is dropped, and the expression loses the $p_{H_2}$ dependence. The values of the reference phases used in this study are: $\Delta G^0_{\text{UH}_3}(298\text{ K}) = -754.52$ meV [24], $E^{DFT}_{\text{U}_3\text{Si}_2} = -45.093$ eV, $E^{DFT}_{\text{UH}_3} = -21.369$ eV and $E^{DFT}_{\alpha\text{-U}} = -9.446$ eV. Note that the use of the DFT energy as a proxy for the free energy of the solid phases introduces a small systematic error, which is cancelled out since the same treatment is applied to all terms of the equation.

# Results and Discussion

## Phase identification

Figure 2 shows the contour plots of all diffraction data for samples $U_3Si_2\_1$ and $U_3Si_2\_2$. With successive 1-d diffraction histograms stacked along the vertical time axis, the colored reflection intensities track the position with respect to $Q = 4\pi \sin\theta / \lambda$ and intensity of diffraction peaks through the experiment.

During the first heating to 350 °C for $U_3Si_2\_1$, all reflections indexed to $U_3Si_2$ structure move to lower Q, resulting from thermal expansion. These reflections continue to shift to lower Q during the isothermal 350 °C holding, which indicates the onset of lattice expansion caused by hydrogen absorption. During the second heating to 375 °C the peaks shift from the combined effects of thermal expansion and hydrogen absorption. Hydrogen absorption is responsible for the peak shift and intensity variation during the 375 °C isothermal period. Meanwhile $UO_2$ peaks begin to increase in intensity. The pattern from the beginning of 375 °C holding period, indicated by the arrow in the steam diagram in Figure 2a, was selected for refining the $U_3Si_2$ hydride structure. The $U_3Si_2$ remains until the end of the 375 °C holding period, when two peaks around 2.4 and 2.7 Å$^{-1}$ form with small diffraction signal and replace the $U_3Si_2$ structure. These two peaks are indexed to 101 and 111 of $U_3Si_5$ and $USi_3$ phase, respectively. The latter's intensity is observed to increase together with the amount of $UO_2$ phase during 400 °C holding and subsequent cooling.

For thermal ramping to 500 °C on sample $U_3Si_2\_2$, thermal expansion causes peak shift to lower Q. During the 10 min isothermal hold at 500 °C, $U_3Si_2$ peaks shift abruptly to lower Q, indicating hydrogen absorption with high reaction rate. This continues until the steam stops, well into the cooling period, an indication that the material is still taking up hydrogen and lattice expansion outweighs the thermal effect would otherwise decrease lattice parameters during cooling.

The formation of $UO_2$ starts at 500 °C. Meanwhile the formation of Si-rich phases, $U_3Si_5$ and $USi_3$, can also be detected from the appearance of 101 ($U_3Si_5$) and 111 ($USi_3$) reflections which overlap with 220, 211, 310 reflections of $U_3Si_2H_x$ phase. These peaks become distinguishable individually on cooling.

## Hydride structure and stability

Figure 3 compares the description of structural models with different hydrogen positions to the diffraction data. Compared with an initial pattern in Figure 3 (a), the pattern at the beginning of 375 °C holding period in Figure 3 (c, d) shows variations in relative peak intensity for 201 and 310 reflections of $U_3Si_2H_x$ phase, caused by absorption of deuterium.

$\chi^2 \cdot R_{F^2}$ results for 12 prospective $^2$H positions combined with 5 stoichiometries are shown in Figure 3 (b). Low values of $\chi^2 \cdot R_{F^2}$ concentrate in two columns, 8*j* (0.032, 0.118, 0.5) and 8*i* (0.202, 0.21, 0), with the two smallest (best) values for $U_3Si_2H_{1.2}$ and $U_3Si_2H_{0.8}$ (1.774 vs. 1.794). Refinement profiles using both these models shown in Figure 3 (c) and (d) are very similar.

The DFT results for $U_3Si_2H_x$, with $x$ varying from 1/32 (dilute conditions) to 2 and H atoms occupying a series of different crystallographic sites are shown in Figure 4 (a,b) for room temperature and 600 K (327 °C). At both temperatures the 8*j* site is favoured for $x \geq 0.5$ and the 2*b* site for $x < 0.5$.

The optimal H absorption site changes with H concentration. H occupying an 8*j* site is unstable in dilute conditions (i.e. $x$ approaching 0) and still has highest (least favourable) formation energy at $x = 0.25$. Shivprasad *et al.* [12] suggested that there are two stages of hydrogen absorption. We propose that this may be connected to H residing at different sites.

At 600 K (327 °C) a similar trend is observed, but all formation energies are lower. The effect of temperature on stability of the hydride phase is shown in Figure 4 (c). Temperature appears to reduce the formation energy of the $U_3Si_2H_x$ less than it reduces the chemical potential of H, resulting in higher overall formation energies. $\Delta G_f$ for the structures with $x \geq 0.5$ and H occupying the 8*j* site are negative up to 500 K (227 °C), with higher H concentrations lowering $\Delta G_f$ ($S^{config}$ is maximal at $x = 2$, as that is when half of the 8*j* sites are occupied). This trend is opposite at temperatures above 500 K due to the entropy of $H_2$, which scales linearly with $x$ and proportionally to $T \ln T$ – see equation (15).

Equation (17) indicates that increasing the partial pressure of hydrogen produces a similar effect to lowering the temperature: the stability of $U_3Si_2H_x$ with $x \geq 0.5$ is increased at all temperatures with increasing $p_{H_2}$. In a steam corrosion environment, where $H_2O$ is dissociated to produce oxides as well hydrides, the thermodynamic equilibria depend also on the oxygen partial pressure, $p_{O_2}$, as well as that of hydrogen. The interplay of the two partial pressures and temperature is non-trivial, and could be investigated by further modelling.

DFT calculations reveal that H absorption at the 8*j* site increases the *a* and *c* lattice parameters. Axial strain is measurably anisotropic, as shown in Figure 4 (d). The unit cell volume increases almost linearly with hydrogen concentration, with ~3.3% volume expansion per formula unit $x$ in $U_3Si_2H_x$. To the extent we can apply a linear relationship to recent experimental unit cell volume results [12], this suggests 4.3% volume expansion per H. This ratio is used below to estimate stoichiometry from lattice strain.

### Structural evolution in situ

Figure 5 shows sequential Rietveld refinement results for sample $U_3Si_2\_1$. Ramping to 350 °C, the lattice strain, shown in Figure 5(a-c) arises purely from thermal expansion. Hydride formation is found to start from about 200 min after exposure to steam during the 350 °C hold as evidenced by the increase in lattice parameters at this constant temperature. For the remaining 140 min hold at 350 °C, the lattice parameters increase by 0.23% *a*, 0.76% *c* and 1.1% volume. By the end of the 350 °C hold, the lattice parameter *a* remains constant for 20 min while *c* still increases. A comparison of the lattice parameters at the start of the first and second thermal ramps, both taken at ~50 °C, shows an expansion in *a*, *c* and *V* of 0.55%, 0.97% and 2.1% respectively due to hydrogen absorption. From the volume expansion defined above, a stoichiometry of $U_3Si_2H_{0.49}$ results from the 350 °C isotherm.

Increase in lattice parameter during the second ramp to 375 °C is due to both thermal expansion and hydrogen absorption, including ongoing increase of lattice volume during an unplanned drop in temperature at 390 min, ~270 °C in Figure 5c. This indicates that once the hydride formation was initiated at 350 °C, uptake continues at lower temperatures.

The lattice expansion between start and end of the 375 °C hold over 30 min is 0.45% *a*, 1.03% *c*, and 1.9% volume. These are greater than at the 350 °C hold for 170 min, hence the reaction is faster at higher temperature, and charges a further 0.45 H to reach an estimated stoichiometry of $U_3Si_2H_{0.94}$ in remaining material. Anisotropic expansion in the lattice is shown by the *a/c* ratio in Figure 5 (c), unlike isotropic thermal expansion in the same temperature range [6].

The Rietveld refinement results for sample $U_3Si_2\_2$ are presented in Figure 6 Initial lattice strain up to 500 °C is thermal expansion. Six min into the 500 °C hold, the lattice parameter *a* remains constant while *c* starts to increase. In another two min, the lattice parameters change abruptly, by 0.54% *a*, 0.43% *c* and 1.5% in volume. The lattice parameter *a* continues to increase until 180 min when the temperature cools to 300 °C, while *c* keeps almost constant after the 500 °C hold and starts to decrease when the temperature drops to 380 °C, at 160 min.

Measured at 70 °C, the lattice parameters are 1.63%, 2.02% and 5.6% larger in *a*, *c* and V respectively, compared with the beginning of the experiment. Based on this volume change, the stoichiometry has now reached $U_3Si_2H_{1.32}$.

Combining the DFT results and the two experimental runs, we propose hydrogen absorption in $U_3Si_2$ takes two steps. From initial heating of sample $U_3Si_2\_1$ to 350 °C (Figure 5c) and DFT results (Figure 4d), the first step is characterized by preferred solution in the 2*b* site, a lattice strain anisotropy of roughly 2:1 in the *c* versus *a* direction, decreasing *a/c* ratio with rising H content, and a volume change of +2.1% until stoichiometry $U_3Si_2H_{0.5}$.

Hydriding at 375 °C showed the same lattice strain anisotropy of 2:1 between *c* and *a* axes up to a stoichiometry of $U_3Si_2H_{0.94}$ and another 1.9% volume strain. This behaviour in lattice strain is matched by the DFT results, which also indicate a change of the preferred 8*j* site.

Further hydriding, which we measured at 500 °C up to $U_3Si_2H_{1.32}$, is characterized by an increasing *a/c* ratio, and a more isotropic lattice strain of roughly 4:3 on *c* and *a* axes. This is seen in the rising *a/c* ratio between 130-180 min in Figure 6c and at higher H-concentration in Figure 4d.

The lattice strain caused by hydriding is compared with Obbard *et al.*'s previous study on the thermal expansion of $U_3Si_2$ in Figure S1 [6]. This offers a way to visualize when hydriding occurs as lattice strain departs from the thermal expansion line. Table S1 lists lattice parameters, strains and stoichiometries relevant to the preceding discussion.

A striking observation from both the contour plots and sequential refinement is hydride formation takes place globally when the temperature is high enough, in our case at 350 °C. We observe no decrease in parent phase peak intensity accompanied by growth in secondary hydride peaks, as expected for a nucleation/growth process. Hydrogen moves rapidly at 500 °C, relative to our temporal resolution of one minute, and lattice expansion of several % affects the whole sample simultaneously.

The observed magnitude and anisotropy in this hydriding lattice strain are respectively hundreds and tens times greater than thermal misfit strains of the order $10^{-4}$ that cause cracking in anisotropic materials [32].

### Reaction mechanism

The total scale factor plotted in Figure 5d is proportional to mole number of crystalline substances in the beam. It decreases steadily until 425 min, half-way into the 375 °C hold, despite $UO_2$ formation during the 350 °C hold. This may be due either to formation of non-crystalline silica or partial loss of sample material due to ejection from the analysis volume. The discontinuity at 373 min is the result of changing H occupancies in the structural model, as stated in the method section. The scale factor after 425 min recovers likely as the result of formation of $UO_2$ offsetting other losses. Based on Figure 5, we propose the following sequence of chemical reactions observed in sample $U_3Si_2\_1$:

350°C: $\qquad U_3Si_2 + \frac{x}{2}H_2O \rightarrow U_3Si_2H_x + \frac{x}{4}O_2$ (or non-crystalline oxide) $\qquad$ (18)

375 °C: $\qquad U_3Si_2H_x + \frac{18}{5}H_2O \rightarrow \frac{9}{5}UO_2 + \frac{2}{5}U_3Si_5 + \left(\frac{18}{5} + \frac{x}{2}\right)H_2$ $\qquad$ (19)

400 °C: $\qquad U_3Si_5 + \frac{8}{3}H_2O \rightarrow \frac{4}{3}UO_2 + \frac{5}{3}USi_3 + \frac{8}{3}H_2$ $\qquad$ (20)

For sample $U_3Si_2\_2$, the $UO_2$ weight fraction rises quickly at 500 °C and replaces the $U_3Si_2H_x$ phase. Si-rich phases appear from the start of the 500 °C hold, while the lattice parameters of the remaining $U_3Si_2$ continue to grow. $UO_2$ phase production is seen to stop at 160 min, which is the time when the steam flow terminates. The constant scale factor from 120 min in Figure 6d would be a combination of oxidation, offset by some sample loss. Sample $U_3Si_2\_2$ at 500 °C experiences a combination of the hydriding and oxidation in equations (18) - (20).

In contrast to the sudden and global changes observed in the $U_3Si_2$ lattice during hydriding, its oxidation to $UO_2$ displays the normal pattern of nucleation and growth, where peak intensity of the $U_3Si_2$ hydride, parent phase is replaced progressively by the oxide, presumably starting from surface and grain boundaries.

Our observations described by equations (18) - (20) are consistent with Yang *et al*'s results [33] and fall in the scope of their thermodynamic predictions under inadequate $H_2O$ content condition. The difference between the temperatures of the formation of $U_3Si_5$ and $USi_3$ phases observed in [33] at 450 and 500 °C, and obtained from our results, at 375 and 400 °C, may possibly be explained by different material forms.

### Post experimental characterization

Post experimental X-ray diffraction on residues from both samples found $U_3Si_5$, $USi_3$ and $UO_2$. To obtain the refinement profile in Figure 7, a Lorentzian broadening parameter was introduced for all three phases while the anisotropic broadening factors [16] were also refined for the $UO_2$ phase, indicating poor crystallinity in the residual material.

Table 1 lists the refinement results from post experiment XRD, and patterns acquired last from in situ NPD. The lattice parameters measured by XRD of $UO_2$ and $USi_3$ are larger, while those of $U_3Si_5$ smaller than literature values [22,34], which implies potential hydrogen absorption or interstitial defects in $UO_2$ and $USi_3$, and Si deficiency in $U_3Si_5$. The lower NPD-measured $UO_2$ fraction at the end of the in situ experiment may be caused by loss of corrosion products from the beam position during reaction, as expected from analysis of the total scale factors, and/or the limited depth of XRD that will be more sensitive to surface phases, such as oxide.

U:Si molar ratios calculated from the total XRD-derived volume fractions are 4.5:1 for sample $U_3Si_2\_1$ and 4.8:1 for sample $U_3Si_2\_2$ – both larger than 3:2 in the original pellets.

Loss of Si from $U_3Si_5$ and apparently insufficient amount of crystalline, Si-rich compounds detected by diffraction are consistent with reports of the production of nano-sized Si [35], non-crystalline silicon oxide [33], and formation of Si from the corrosion process [36]. The confirmation of Si-rich phases by both in situ NPD and ex situ XRD match progressive corrosion mechanisms where Si dissociates from $U_3Si_2$ and migrates inwards to form Si-rich phase [33]. U reacts with O from water molecules undergoing release hydrogen [37].

|  | $UO_2$ | $U_3Si_5$ | $USi_3$ | $U_3Si_2$ |
|---|---|---|---|---|
| Reference lattice parameters (Å) | 5.468 | $a$=4.069, $c$=3.843 | 4.034 | |
| **$U_3Si_2$_1:** | | | | |
| XRD lattice parameters (Å) | 5.4773(5) | $a$=4.03(1), $c$=3.80(1) | 4.052(1) | |
| XRD vol. fraction (%) | 86.4(5) | 6.4(5) | 7.1(2) | |
| NPD vol. fraction (%) | 67(4) | | 33(4) | |
| **$U_3Si_2$_2:** | | | | |
| XRD lattice parameters (Å) | 5.4742(3) | $a$=4.067(5), $c$=3.806(8) | 4.056(4) | |
| XRD vol. fraction (%) | 86.3(3) | 5.6(3) | 8.1(1) | |
| NPD vol. fraction (%) | 62.9(22) | 10.2(14) | 15.9(13) | 11(1) |

Table 1: Lattice parameters and volume fractions for post experiment samples obtained from XRD, from the literature [22,34], and from last-measured in situ NPD patterns.

In scanning electron microscopy (SEM), sample $U_3Si_2$_1 consists of loosely connected particles full of cracks shown in Figure 8 (a). Backscattered electron images on the polished sample (b) show striated, layered microstructure with distinctive mass densities as shown in (c), named as light bulk phase (LBP) and dark bulk phase (DBP). Their isolated analogue shown in (d) are named as dark isolated phase (DIP) and light isolated phase (LIP). The striated morphologies are like those observed by Sooby Wood *et al* [10].

Subject to caution in interpreting un-calibrated EDS, the chemical composition of the dark phases DBP and DIP may be interpreted as relatively silicon-rich $USi_x$ + $SiO_2$ while those of the light phases LBP and LIP as uranium-rich $USi_x$ +$UO_2$. The detailed EDS results are listed in Table S2. $U_3Si_2$ thus corrodes in steam to a mixture of $UO_2$ + $SiO_2$ + silicon-rich uranium compounds.

From the SEM topographies of sample $U_3Si_2$_2 such as Figure 8 (e), significant numbers of vent holes were observed besides the prevalent cracks running through all particles. Based on Eq. (12), these would logically result from the venting of $H_2$ that can no longer be absorbed by hydriding.

Backscattered electron images on the polished sample show striated phases, DBP and LBP, growing close to each other as shown in Figure 8 (h). The phase composition of DBP and LBP can again be interpreted as $SiO_2$ + $USi_x$ and $UO_2$ + $USi_x$. But different from sample $U_3Si_2$_1, high contrast in grain boundary areas is observed and pointed out by arrows in Figure 8 (g). The corroded grain boundaries such as highlighted in Figure 8 (g) are close to $UO_2$ in composition. This is the sign of intergranular corrosion that starts from the grain boundaries composed of mainly uranium oxide, which can be seen directly in an elemental mapping in Figure S2. The much darker DIP found in $U_3Si_2$_2 is approximately pure $SiO_2$, according to the EDS analysis.

These quenched cross section pictures suggest that the corrosion reaction besides the hydride formation is driven in at least two ways simultaneously: pushing Si into the bulk material while forming uranium oxide in the grain boundaries [33]; and segregation of Si interleaved striations causing separation of U and Si concentration. Both routes are related to the migration/diffusion of Si and O atoms with respect to the relatively heavy U atoms.

## Conclusions

DFT modelling, coupled with a rigorous thermodynamic description of hydrogen chemical potential, resolves a transition in preferred site for H that takes place at a stoichiometry of $U_3Si_2H_{0.5}$ from 2*b* to 8*j* in $U_3Si_2$, with increasing H content. This was confirmed using in situ neutron diffraction, where Rietveld refinement showed H is accommodated at the 8*j* interstitial site in $U_3Si_2$, as it undergoes corrosion in deuterated steam, and before the sample is itself consumed by oxidation. This further demonstrates the use of neutron diffraction for in situ experiments on nuclear materials.

The investigation has offered a view of the transient phases and reaction steps taking place at four different temperatures. At 350 °C, $U_3Si_2$ transforms into hydride, with likely byproducts of poorly crystalline oxide or free oxygen. At 375 °C, the hydride phase gradually decomposes into $UO_2$ and $U_3Si_5$ that is sub-stoichiometric in Si. At 400 °C and above, $U_3Si_5$ oxidizes to $UO_2$ and $USi_3$. Based on observed final microstructures and molar ratios from XRD, unaccounted-for Si content likely resides in finely distributed amorphous silicon or/and silicates.

These reactions take place progressively, by nucleation and growth of second phases at surface or grain boundaries in already-hydrided $U_3Si_2H_x$. On the scale of our samples, and the time resolution of our experiment, hydriding of $U_3Si_2$ is observed globally, affecting all $U_3Si_2$ in our sample simultaneously. It takes place with increasing rate, to higher H stoichiometries, as we raise the temperature.

$U_3Si_2$ fragmentation caused by hydriding can be explained by the strong anisotropy in the lattice expansion, which creates a linear differential strain of ~0.5% in the *a* and *c* axes even in initial stages of hydriding at 350 °C – enough to crack apart a polycrystalline aggregate that may have withstood an isotropic expansion in its grains. This was corroborated by ensemble averages of a large number of DFT simulations, which show that the single-crystal structural evolution upon hydriding is responsible for linear increase in volumetric strain accompanied by non-linear changes in a/c ratio. The trend depends primarily on hydrogen content and is insensitive to both temperature and hydrogen partial pressure.

## Acknowledgement

This research was undertaken with the assistance of resources and services from the National Computational Infrastructure, which is supported by the Australian Government; through the UNSW-NCI partner trial scheme; the Multi-modal Australian ScienceS Imaging and Visualisation Environment (MASSIVE); the Pawsey Supercomputing Centre, which is supported by the Australian Government and the Government of Western Australia; and was enabled by Intersect Australia Limited. The authors are grateful for beamtime allocated at the Australian Centre for neutron scattering under proposal P8186, and thank Grant Griffiths, John Macleod and Richard Collins for assistance with sample handling.

## Data availability

The raw/processed data required to reproduce these findings cannot be shared at this time due to technical or time limitations.

## Reference

[1] J.T. White, A.T. Nelson, J.T. Dunwoody, D.J. Safarik, K.J. McClellan, Corrigendum to "Thermophysical properties of $U_3Si_2$ to 1773 K" [J. Nucl. Mater. (2015) 464 (275–280)] (S002231151500241X)(10.1016/j.jnucmat.2015.04.031), Journal of Nuclear Materials. 484 (2017) 386–387. https://doi.org/10.1016/j.jnucmat.2016.11.015.

[2] M.R. Finlay, G.L. Hofman, J.L. Snelgrove, Irradiation behaviour of uranium silicide compounds, Journal of Nuclear Materials. 325 (2004) 118–128. https://doi.org/10.1016/j.jnucmat.2003.11.009.

[3]   K. Sung-Joong, the Opal (Open Pool Australian Light-Water) Reactor in Australia, Nuclear Engineering and Technology. 38 (2006) 443–448.

[4]   A. Gonzales, J.K. Watkins, A.R. Wagner, B.J. Jaques, E.S. Sooby, Challenges and opportunities to alloyed and composite fuel architectures to mitigate high uranium density fuel oxidation: uranium silicide, Journal of Nuclear Materials. (2021) 153026. https://doi.org/10.1016/j.jnucmat.2021.153026.

[5]   A. Mohamad, Y. Ohishi, H. Muta, K. Kurosaki, S. Yamanaka, Thermal and mechanical properties of polycrystalline U 3 Si 2 synthesized by spark plasma sintering, (2018). https://doi.org/10.1080/00223131.2018.1480431.

[6]   E.G. Obbard, K.D. Johnson, P.A. Burr, D.A. Lopes, D.J. Gregg, K.D. Liss, G. Griffiths, N. Scales, S.C. Middleburgh, Anisotropy in the thermal expansion of uranium silicide measured by neutron diffraction, Journal of Nuclear Materials. 508 (2018) 516–520. https://doi.org/10.1016/j.jnucmat.2018.04.049.

[7]   T.L. Ulrich, S.C. Vogel, J.T. White, D.A. Andersson, E. Sooby Wood, T.M. Besmann, High temperature neutron diffraction investigation of U3Si2, Materialia. 9 (2020) 100580. https://doi.org/10.1016/j.mtla.2019.100580.

[8]   Y. Miao, K.A. Gamble, D. Andersson, B. Ye, Z.G. Mei, G. Hofman, A.M. Yacout, Gaseous swelling of $U_3Si_2$ during steady-state LWR operation: A rate theory investigation, Nuclear Engineering and Design. 322 (2017) 336–344. https://doi.org/10.1016/j.nucengdes.2017.07.008.

[9]   S.C. Middleburgh, R.W. Grimes, E.J. Lahoda, C.R. Stanek, D.A. Andersson, Non-stoichiometry in U3Si2, Journal of Nuclear Materials. 482 (2016) 300–305. https://doi.org/10.1016/j.jnucmat.2016.10.016.

[10]  E. Sooby Wood, J.T. White, C.J. Grote, A.T. Nelson, $U_3Si_2$ behavior in $H_2O$: Part I, flowing steam and the effect of hydrogen, Journal of Nuclear Materials. 501 (2018) 404–412. https://doi.org/10.1016/j.jnucmat.2018.01.002.

[11]  S.C. Middleburgh, A. Claisse, D.A. Andersson, R.W. Grimes, P. Olsson, S. Mašková, Solution of hydrogen in accident tolerant fuel candidate material: U3Si2, Journal of Nuclear Materials. 501 (2018) 234–237. https://doi.org/10.1016/j.jnucmat.2018.01.018.

[12]  A.P. Shivprasad, V. Kocevski, T.L. Ulrich, J.R. Wermer, D.A. Andersson, J.T. White, The $U_3Si_2$ - H system, Journal of Nuclear Materials. 558 (2022) 153278. https://doi.org/10.1016/j.jnucmat.2021.153278.

[13]  A.J. Studer, M.E. Hagen, T.J. Noakes, Wombat: The high-intensity powder diffractometer at the OPAL reactor, Physica B: Condensed Matter. 385–386 (2006) 1013–1015. https://doi.org/10.1016/j.physb.2006.05.323.

[14]  D.R. Black, D. Windover, A. Henins, J. Filliben, J.P. Cline, Certification of Standard Reference Material 660B, Powder Diffraction. 26 (2011) 155–158. https://doi.org/10.1154/1.3591064.

[15]  H.M. Rietveld, A profile refinement method for nuclear and magnetic structures, J Appl. Crystallogr. 2 (1969) 65–71. https://doi.org/10.1107/S0021889869006558.

[16]  A. C. Larson, R B Von Dreele, General Structure Analysis System (GSAS), 2004.

[17]   S.C. Vogel, Gsaslanguage: A GSAS script language for automated Rietveld refinements of diffraction data, Journal of Applied Crystallography. 44 (2011) 873–877. https://doi.org/10.1107/S0021889811023181.

[18]   J.C. Slater, Atomic radii in crystals, The Journal of Chemical Physics. 41 (1964) 3199–3204. https://doi.org/10.1063/1.1725697.

[19]   R.N.R. Mulford, F.H. Ellinger, W.H. Zachariasen, A New Form of Uranium Hydride, Journal of the American Chemical Society. 76 (1954) 297–298.

[20]   R.E. Rundle, The Hydrogen Positions in Uranium Hydride by Neutron Diffraction, Journal of the American Chemical Society. 73 (1951) 4172–4174. https://doi.org/10.1021/ja01153a035.

[21]   D.R.J. Boyd, Infrared spectrum of trideuterosilane and the structure of the silane molecule, The Journal of Chemical Physics. 23 (1955) 922–926. https://doi.org/10.1063/1.1742148.

[22]   T. Miyadai, H. Mori, T. Oguchi, Y. Tazuke, H. Amitsuka, T. Kuwai, Y. Miyako, Magnetic and electrical properties of the U-Si system (part II), Journal of Magnetism and Magnetic Materials. 104–107 (1992) 47–48. https://doi.org/10.1016/0304-8853(92)90697-M.

[23]   M.W. Finnis, A.Y. Lozovoi, A. Alavi, THE OXIDATION OF NIAL: What Can We Learn from Ab Initio Calculations?, Annual Review of Materials Research. 35 (2005) 167–207. https://doi.org/10.1146/annurev.matsci.35.101503.091652.

[24]   W.M. Haynes, D.R. Lide, T.J. Bruno, CRC handbook of chemistry and physics: a ready-reference book of chemical and physical data., 97th ed., CRC Press, Boca Raton, Florida, 2016.

[25]   G. Kresse, J. Furthmuller, Efficient iterative schemes for ab initio total-energy calculations using a plane-wave basis set, Phys. Rev. B. 54 (1996) 11169. http://dx.doi.org/10.1103/PhysRevB.54.11169.

[26]   D. Joubert, From ultrasoft pseudopotentials to the projector augmented-wave method, Physical Review B - Condensed Matter and Materials Physics. 59 (1999) 1758–1775. https://doi.org/10.1103/PhysRevB.59.1758.

[27]   J.P. Perdew, K. Burke, M. Ernzerhof, Generalized gradient approximation made simple, Phys. Rev. Lett. 77 (1996) 3865. http://dx.doi.org/10.1103/PhysRevLett.77.3865.

[28]   S. Dudarev, G. Botton, Electron-energy-loss spectra and the structural stability of nickel oxide: An LSDA+U study, Physical Review B - Condensed Matter and Materials Physics. 57 (1998) 1505–1509. https://doi.org/10.1103/PhysRevB.57.1505.

[29]   M.J. Noordhoek, T.M. Besmann, D. Andersson, S.C. Middleburgh, A. Chernatynskiy, Phase equilibria in the U-Si system from first-principles calculations, Journal of Nuclear Materials. 479 (2016) 216–223. https://doi.org/10.1016/j.jnucmat.2016.07.006.

[30]   P.A. Burr, M.W.D. Cooper, Importance of elastic finite-size effects: Neutral defects in ionic compounds, Physical Review B. 96 (2017) 094107. https://doi.org/10.1103/PhysRevB.96.094107.

[31]   R. Grau-Crespo, S. Hamad, C.R.A. Catlow, N.H. de Leeuw, Symmetry-adapted configurational modelling of fractional site occupancy in solids, Journal of Physics Condensed Matter. 19 (2007). https://doi.org/10.1088/0953-8984/19/25/256201.

[32] A.A. Rezwan, A.M. Jokisaari, M.R. Tonks, Modeling brittle fracture due to anisotropic thermal expansion in polycrystalline materials, Computational Materials Science. 194 (2021) 110407. https://doi.org/10.1016/j.commatsci.2021.110407.

[33] J.H. Yang, D.S. Kim, D.J. Kim, S. Kim, J.H. Yoon, H.S. Lee, Y.H. Koo, K.W. Song, Oxidation and phase separation of $U_3Si_2$ nuclear fuel in high-temperature steam environments, Journal of Nuclear Materials. 542 (2020). https://doi.org/10.1016/j.jnucmat.2020.152517.

[34] R.W.G. Wyckoff, Crystal Structures, second edition, Interscience Publishers, 1963.

[35] R.W. Harrison, C. Gasparrini, R.N. Worth, J. Buckley, M.R. Wenman, T. Abram, On the oxidation mechanism of U 3 Si 2 accident tolerant nuclear fuel, Corrosion Science. 17 (2020) 108822. https://doi.org/10.1016/j.corsci.2020.108822.

[36] E. Sooby Wood, C. Moczygemba, G. Robles, Z. Acosta, B.A. Brigham, C.J. Grote, K.E. Metzger, L. Cai, High temperature steam oxidation dynamics of $U_3Si_2$ with alloying additions: Al, Cr, and Y, Journal of Nuclear Materials. 533 (2020). https://doi.org/10.1016/j.jnucmat.2020.152072.

[37] E. Jossou, L. Malakkal, N.Y. Dzade, A. Claisse, B. Szpunar, J. Szpunar, DFT + U Study of the adsorption and dissociation of water on clean, defective, and oxygen-Covered U3Si2{001}, {110}, and {111} Surfaces, Journal of Physical Chemistry C. 123 (2019) 19453–19467. https://doi.org/10.1021/acs.jpcc.9b03076.

# Figures

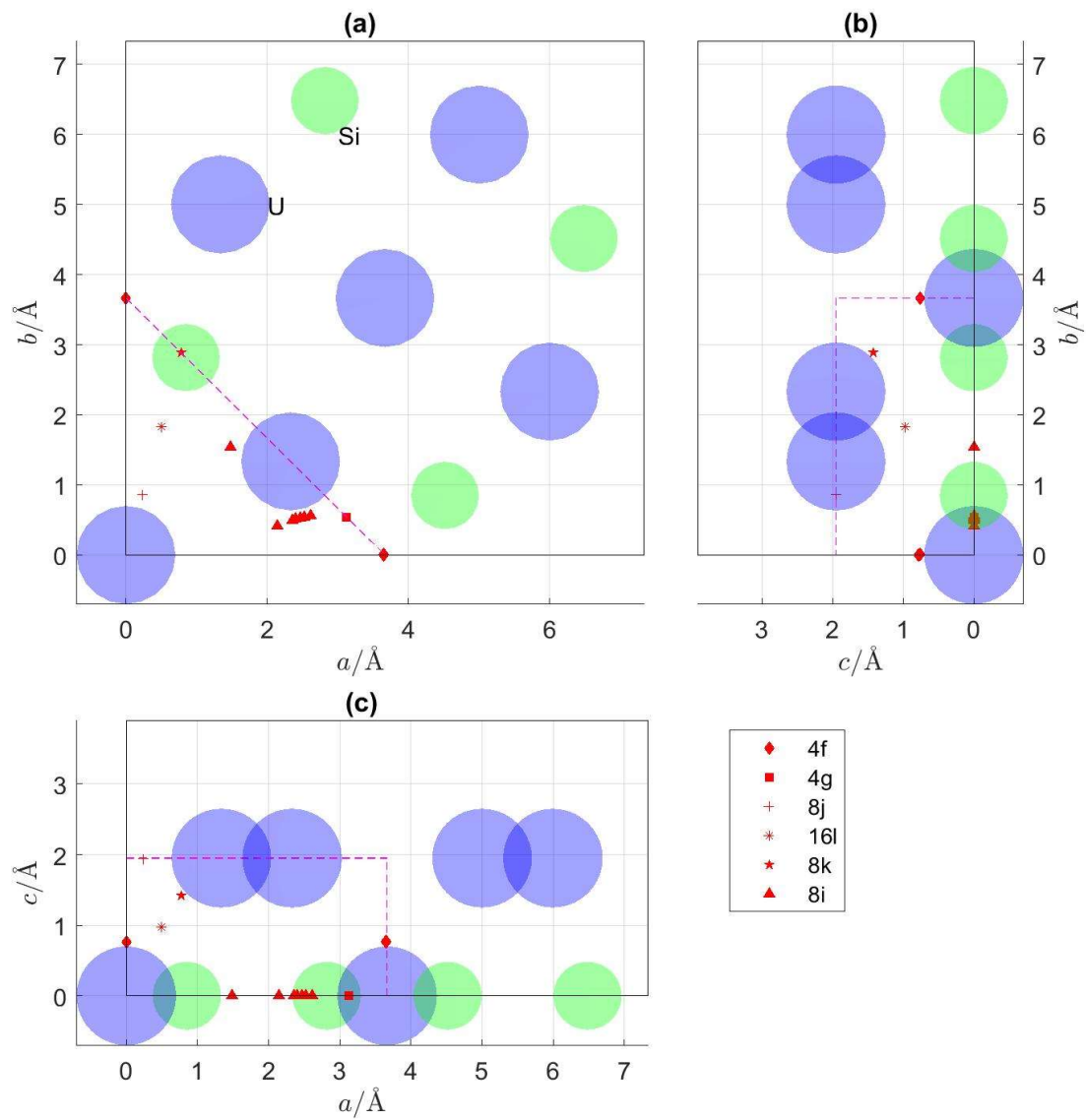

**Figure 1:** (a)-(c) The room temperature U$_3$Si$_2$ unit cell (excluding repeated atoms) viewed parallel to c-, a- and b- axis with 12 candidate H positions pointed out in the irreducible volume, 1/16 of the volume of the unit cell (dashed volume). Shaded circles show U (large, blue) and Si (small, green) atoms.

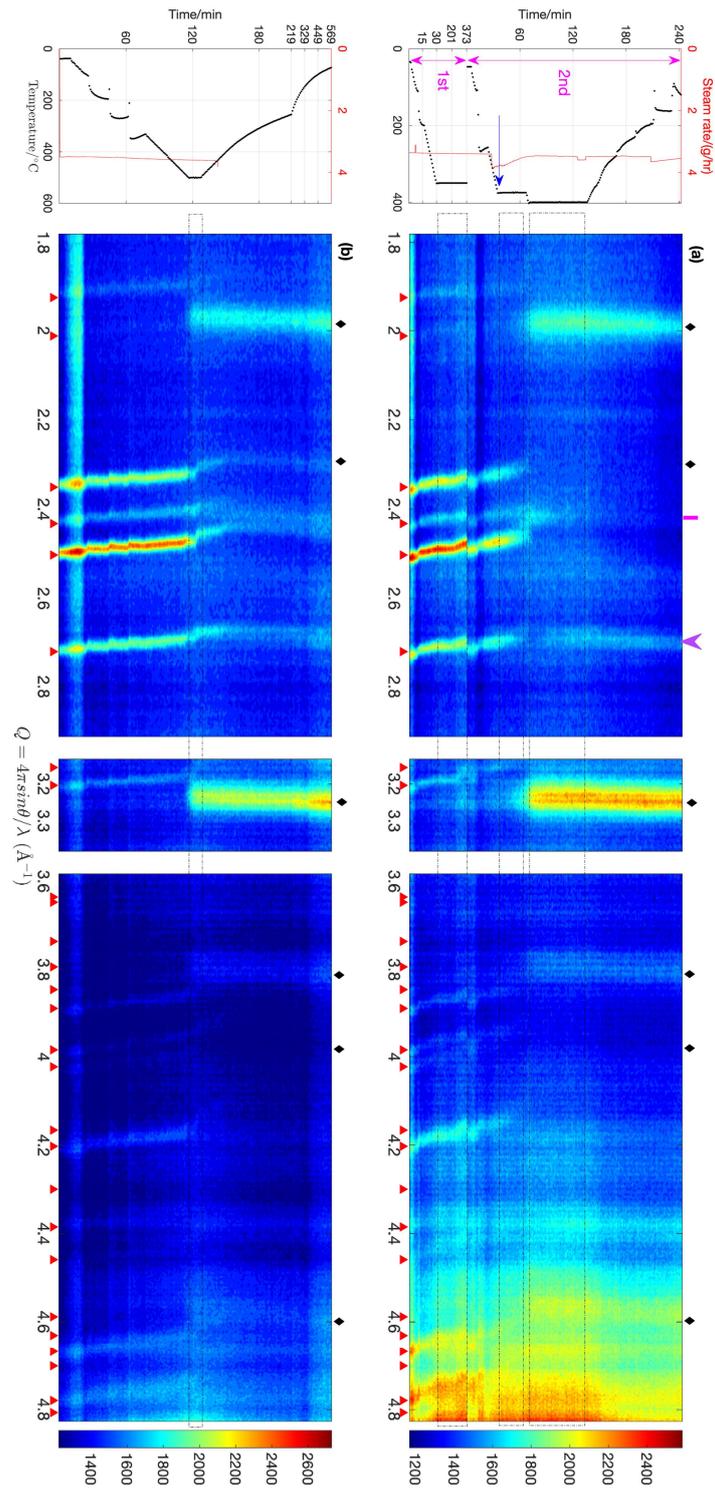

**Figure 2:** Steam and heating condition (left) and contour plot of diffraction data (right) excluding stainless-steel peaks for the sample U$_3$Si$_2$_1 (a) and U$_3$Si$_2$_2 (b). Phases are indexed with different symbols: U$_3$Si$_2$ (▲), UO$_2$ (♦), USi$_3$ (►), U$_3$Si$_5$ (|). Only one reflection from each of USi$_3$ (111 reflection) and U$_3$Si$_5$ (101 reflection) phase is identified. The isothermal regions where significant peak shifting happens are annotated within dashed-line rectangles. The first and second cycle of heating are pointed out for U$_3$Si$_2$_1. An arrow (→) indicates the beginning of 375 °C hold during the second thermal cycle on U$_3$Si$_2$_1.

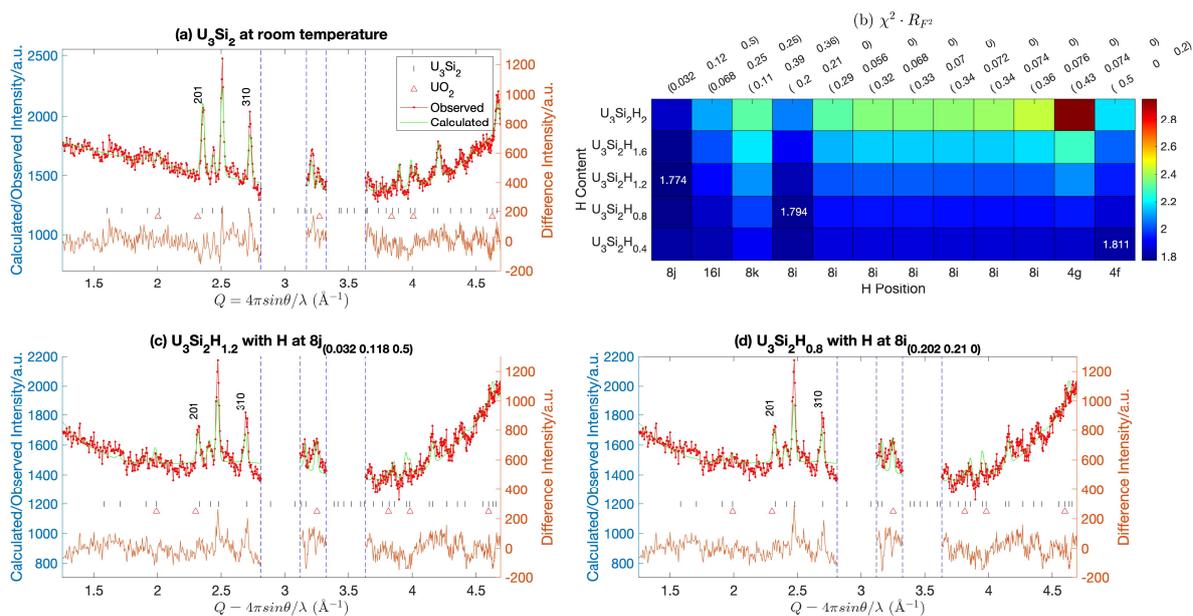

**Figure 3:** (a) Rietveld refinement profile for a model containing $UO_2$ and $U_3Si_2$ phase against initial NPD data at room temperature and before exposure to steam; (b) Refinement criteria $\chi^2 \cdot R_{F^2}$ values for data shown in a) for all $^2$H-content $U_3Si_2H_{0.4-2}$ (vertical axis) models with 12 candidate $^2$H positions (horizontal axis); lowest values for 8*j*, 8*i* and 4*f* sites are annotated; atomic coordinates and sites are labelled at the top and bottom x-axis, respectively; (c) (d) Rietveld refinement profiles using structural models with $UO_2$ and $U_3Si_2H_{1.2}$ phases with $^2$H at 8*j* (0.032, 0.118, 0.5) and $U_3Si_2H_{0.8}$ with $^2$H at 8*i* (0.202, 0.21, 0) positions, shown against in situ neutron diffraction data at the beginning of the 375 °C hold. Two strong peaks from the sample holder are excluded.

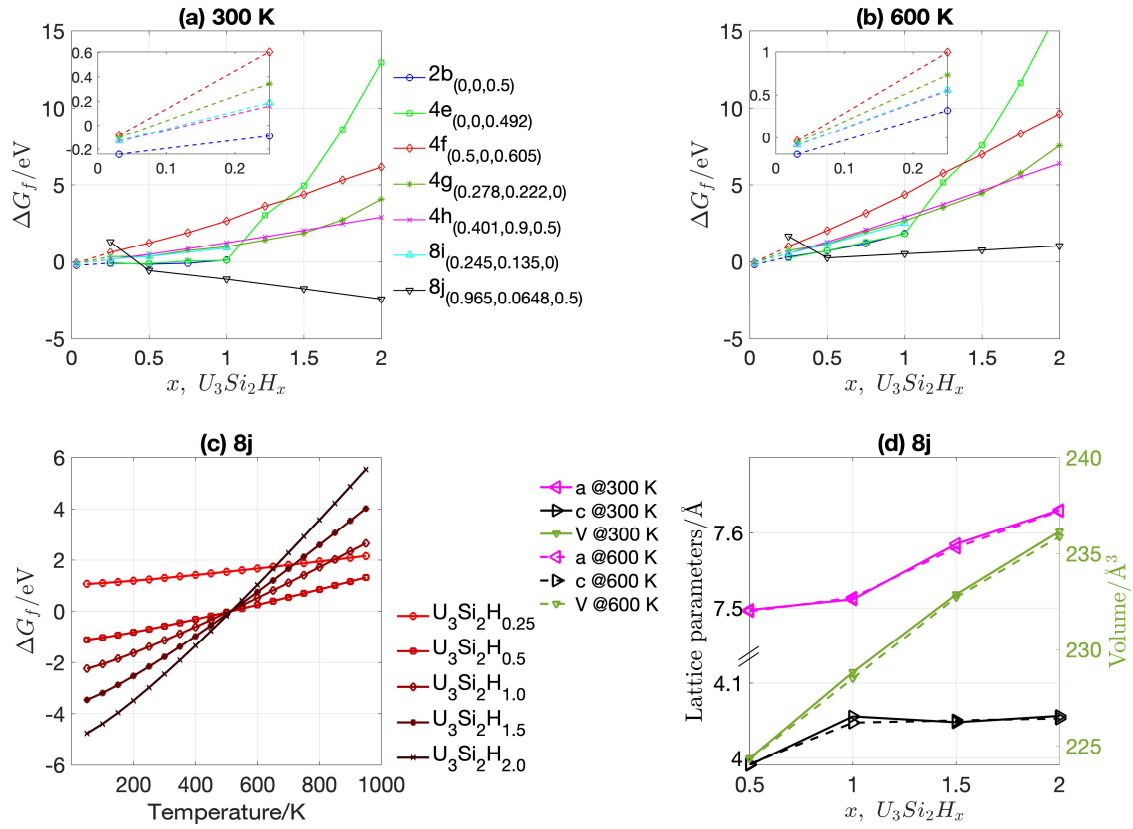

**Figure 4**: Calculated free energy of formation of $U_3Si_2H_x$ at standard $H_2$ partial pressure with different amount of H occupying crystallographic sites at (a) 300 K and (b) 600 K. All structures comprise a 1x1x2 supercell, except x = 1/32 comprising a 2x2x4 supercell; (c) Calculated free energy of formation for the $U_3Si_2H_x$ at standard $H_2$ partial pressure with H only occupying 8*j* site, over a temperauture range; (d) Calculated lattice parameters and unit cell volumes for $U_3Si_2H_x$ structures with H occupying the 8*j* site at 300 K and 600 K. Lines through points are guides to the eye.

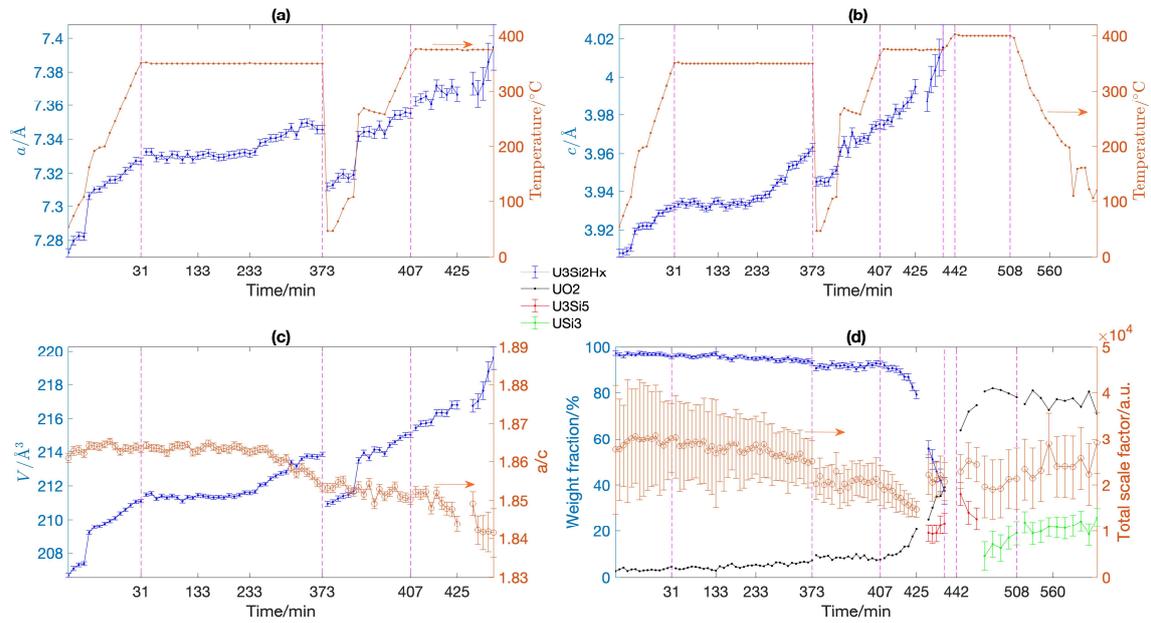

**Figure 5:** Unit cell and phase evolution of sample $U_3Si_2\_1$ obtained from sequential Rietveld refinement using in situ temperature dependent NPD data: (a) (b) lattice parameters, and temperature, (c) volume (dot-lines) and a/c ratio (circle-lines), (d) weight fractions for all phases (dot-lines) and total scale factors (circle-lines). Sections of data when the number of phases start to change are omitted due to the instability of refinements. The dashed-vertical lines are located at the beginning/ending of the isothermal holding periods. Lines through points are a guide to the eye.

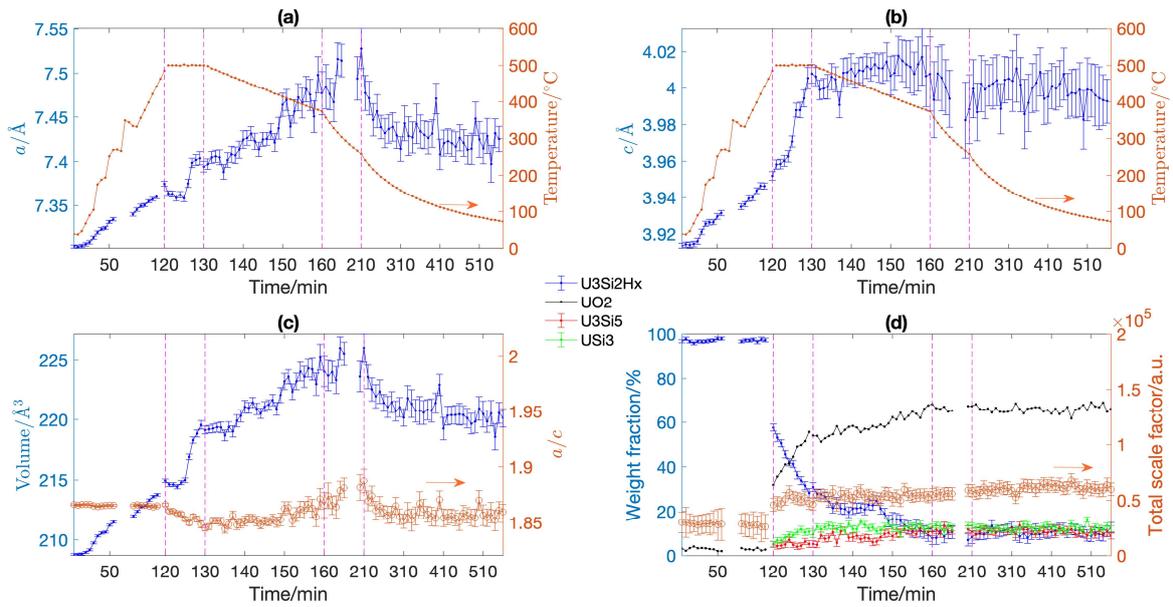

**Figure 6:** Unit cell and phase evolution of sample U$_3$Si$_2$_2 from sequential Rietveld refinement using in situ temperature dependent NPD data: (a) (b) U$_3$Si$_2$ lattice parameters, and temperature, (c) U$_3$Si$_2$ volume (dot-lines) and a/c ratio (circle-lines), (d) weight fractions for all phases refined (dot-lines) and total scale factors (circle-lines). The dashed-vertical lines help to read the irregular x-axis tick labels and lines through points are a guide to the eye.

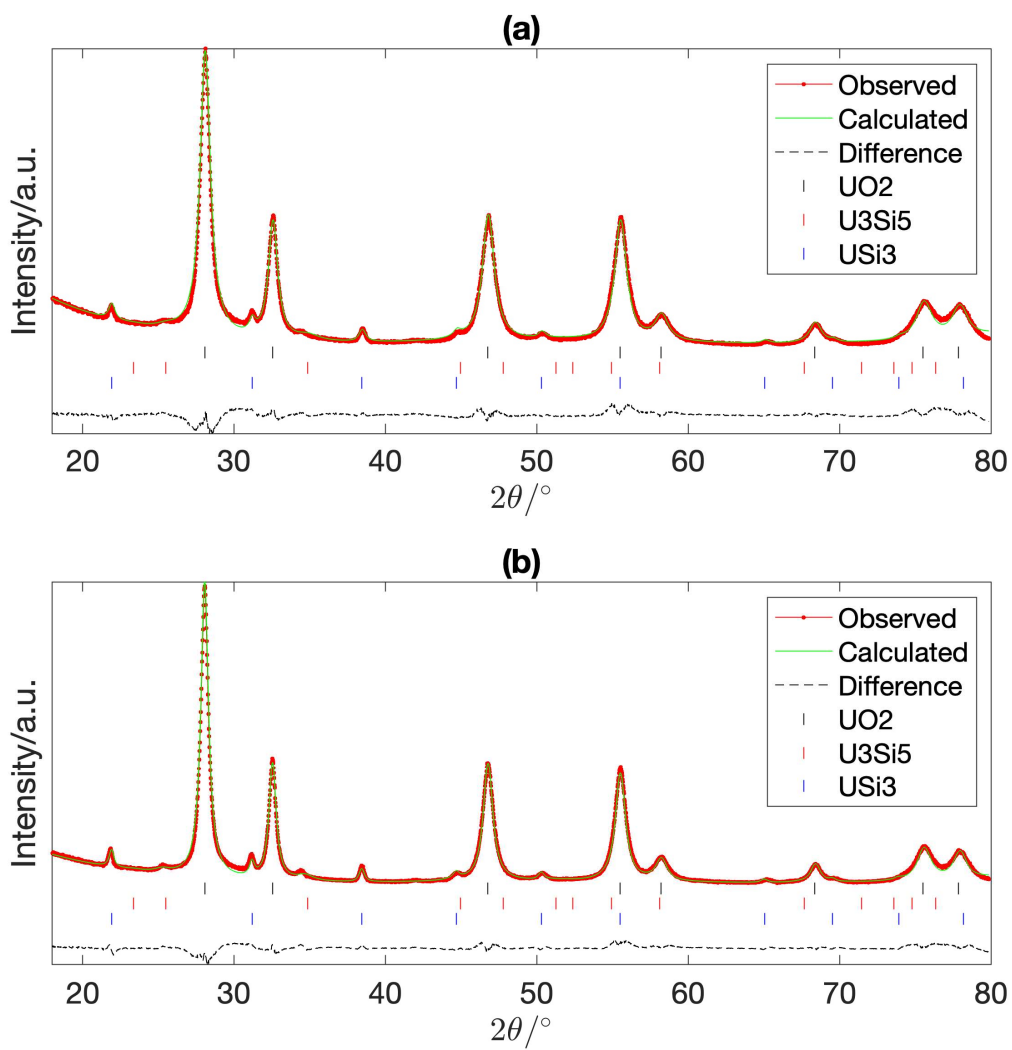

**Figure 7:** XRD data and Rietveld refinement profiles for (a) $U_3Si_2\_1$ and (b) $U_3Si_2\_2$ after in situ experiments using three phases $UO_2$ (*Fm$\bar{3}$m*), $U_3Si_5$ (*P6/mmm*) and $USi_3$ (*Pm$\bar{3}$m*), with refinement goodness of fit $\chi^2$ 16.04 and 16.65 respectively.

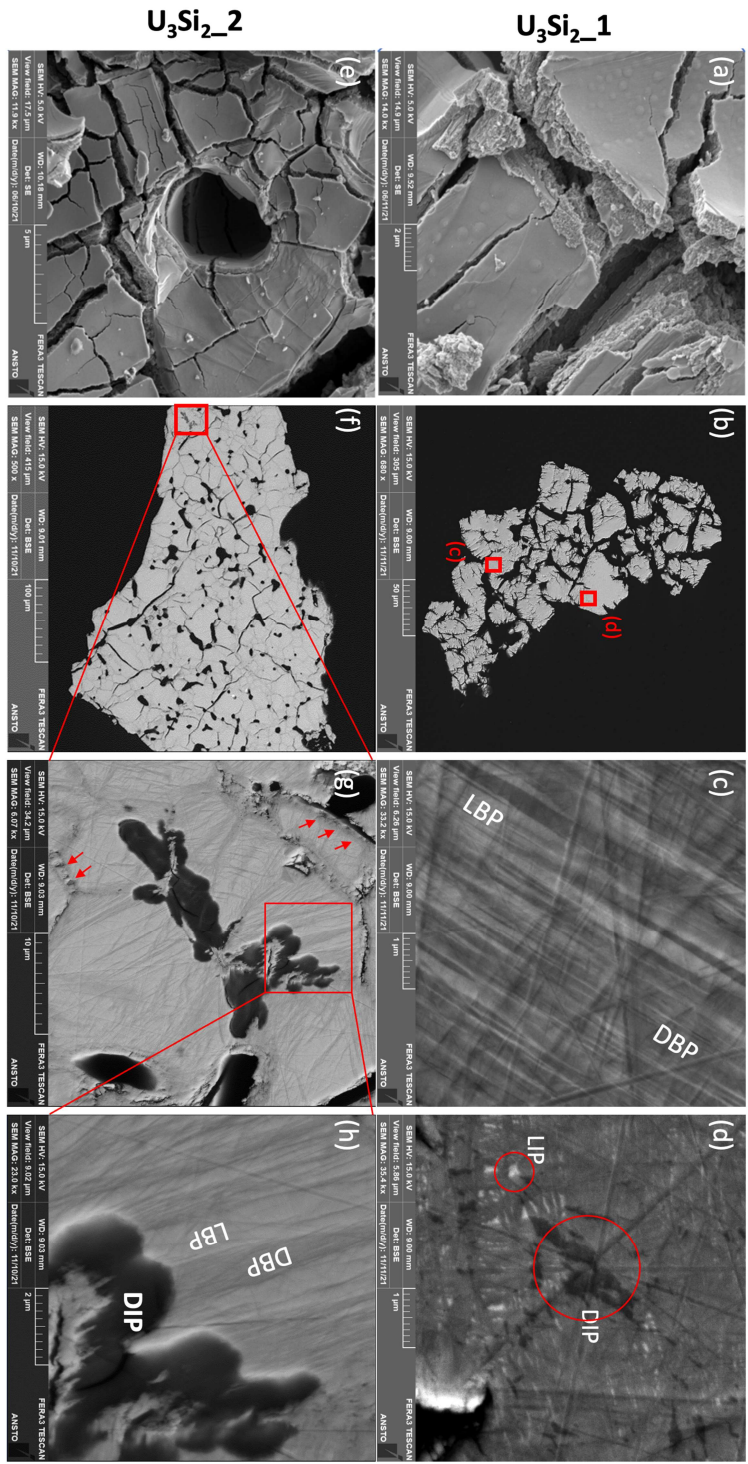

**Figure 8:** Secondary (a, e) and backscattering electron images (b-d), (f-h) of the sample U$_3$Si$_2$_1 (a-d) and U$_3$Si$_2$_2 (e-h) after in situ corrosion experiments. (c) and (d) are zoom images of red rectangles in (b); (g) and (h) are progressively zoomed images of red rectangle in (f). Annotations in (c), (d) and (h) denote light bulk phase (LBP), dark bulk phase (DBP), dark isolated phase (DIP) and light isolated phase (LIP). Arrows in (g) emphasise a grain boundary.